\documentclass[10pt,letterpaper]{article}
\usepackage{opex3}
 \usepackage{graphicx}
\usepackage{dcolumn}
\usepackage{bm}
\usepackage{array}
\usepackage{amsmath}
\usepackage{multirow}
\usepackage{cite}

\newlength{\figwidth} 

\newcommand{\beq}{\begin{equation}}

\newcommand{\beql}[1]{\begin{equation}\label{#1}}

\newcommand{\eeq}{\end{equation}}
\newcommand{\bsp}{\begin{split}}
\newcommand{\esp}{\end{split}}

\newcommand{\Eq}[1]{Eq.~(\ref{#1})}

\begin{document}
\title{Fourier finite element modeling of light emission in waveguides: 2.5-dimensional FEM approach}

\author{ Yangxin Ou,$^{1}$ David Pardo,$^{2,3,4}$ and Yuntian Chen$^{1,5}$$^*$}

\address{$^1$School of Optical and Electronic Information, Huazhong University of Science and Technology, Wuhan, China\\
$^2$Dpto. de Matem\'{a}tica Aplicada y Estad\'{i}stica e Investigaci\'{o}ón Operativa, Barrio Sarriena S/N, Facultad de Ciencia y Tecnolog\'{i}a, 48940 Leioa (Bizkaia), Spain\\$^3$BCAM (Basque Center for Applied Mathematics), Bilbao, Spain\\$^4$Ikerbasque (Basque Foundation for Sciences), Bilbao, Spain\\$^5$Wuhan National Laboratory of Optoelectronics, Huazhong University of Science and Technology, Wuhan, China}

\email{$^*$yuntian@hust.edu.cn}

\begin{abstract}
We present a Fourier finite element modeling of light emission of dipolar emitters  coupled to infinitely long waveguides. Due to the translational symmetry, the three-dimensional (3D) coupled waveguide-emitter system can be decomposed into a series of independent 2D problems (2.5D), which reduces the computational cost. Moreover, the reduced 2D problems can be extremely accurate, compared to its 3D counterpart. Our method can precisely quantify the total emission rates, as well as the fraction of  emission rates into different modal channels for waveguides with  arbitrary cross-sections. We compare our method  with dyadic Green's function for the light emission in single mode metallic nanowire, which  yields an excellent agreement. This method is applied in multi-mode waveguides, as well as multi-core waveguides. We further show that our method has the full capability of including dipole orientations, as illustrated via a rotating dipole, which leads to unidirectional excitation of guide modes. The 2.5D Finite Element Method (FEM) approach proposed here can be applied for various waveguides, thus it is useful to interface single-photon single-emitter in nano-structures, as well as for other scenarios involving coupled waveguide-emitters.
\end{abstract}


\section{Introduction}

Controllably funneling emitted photons into certain modes  has potential applications in developing single photon sources, low threshold lasers, optical nanoantennas,  ultra-fast optical switches and many others \cite{Lodahl2015,Frimmer2011,Lund2008,tonglimin2013,Femius2006,Hummer2013,gongqihuang2014,Kneipp1997,Chen2009,Ulrik2013OL,Søndergaard2001,Falk2009,Crozier2003,Christian2002}. This imposes experimental challenges in controlling light emitters and waveguides, as well as theoretical difficulties in predicting light emission into complex photonic environment \cite{Vos2004,Colas2010,Chang2007,Hecker1999}.  There are considerable efforts for controlling the spontaneous  emission (SE) using quantum dots \cite{perter2008},  molecules \cite{V.sandoghdar2006}, NV-centers \cite{Ulrik2013} in photonic crystal platforms \cite{Chang2006,
Arcari2014}, plasmonic waveguides, nanoantennas, leading to suppressed or enhanced  emission rate, or directional emission in various photonic environments. Meanwhile, different theoretical tools \cite{PRB,Niles2009,Chen2010,Pardo,Jakob2013}  have been developed to study the mode properties, and light emission  in different photonic structures.

In the filed of waveguides, Chen et al. \cite{PRB} studied  emission rates $\gamma$ of waveguide modes,  using 2D mode profiles by taking advantage of  the translation symmetry of the waveguide. However, the presence of the dipolar emitter breaks the translation symmetry of the solution, which  gives rise to difficulty in calculating the total emission rate $\gamma_{total}$. Thus, in order to obtain $\gamma_{total}$, or the fraction ( $\gamma_{m}/\gamma_{total}$) of emitted photon into a specific waveguide mode, the authors in \cite{PRB} further developed a 3D FEM to study the total emission rate, consisting of bound modes, radiation modes, and other non-radiative channels. Chen's 3D FEM model, though equipped with non-trivial mode matching technique, is  limited by the fact that only one dominatingly guided mode is allowed to be excited. However, multi-mode excitation is needed as the cross-section of the waveguide becomes large, or  when  random orientated dipoles excite two different polarization modes in an elliptical waveguide \cite{ellipse}. We refer to Ref.\cite{PRB} for the motivation of our choice of choosing FEM as our method to study the coupled waveguide-emitter system.

In this paper, we lift the  restriction of the single dominating mode excitation  in \cite{PRB} by developing a 2.5D FEM approach. The aforementioned difficulty can be solved by fully taking advantage of the point nature of the dipolar emitter as well as the 2D profile of material indices, meaning that the coupling between the waveguide modes and the emitter, which are associated with different propagation constants (${{k_{nz}}}$), turn out to be decoupled. This leads to a Fourier finite element implementation of light emission from point dipoles in waveguides. Essentially, the 3D coupled waveguide-emitter problem is decomposed into a series of 2D uncoupled waveguide-emitter problems, each of which is associated with a single ${{k_{nz}}}$. Our method resembles  a spectrum presentation of the dyadic Green's function of the waveguide in terms of ${{k_{nz}}}$. The 2.5D FEM approach can handle multi-mode excitation for waveguide with arbitrary cross-section, in which the dipole orientation can be arbitrary as well.

The paper is organized as follows. Section 2 describes the foundation of a 2.5D FEM approach, and the basic theory of light emission. In Section 3, we first provide an example of a single metallic nanowire as a benchmark of the 2.5 FEM approach  against the Green's function techniques. We further apply our 2.5D FEM approach to multi-mode  and a multi-core waveguide. Finally, we show that the 2.5D FEM approach can be used to realize unidirectional excitation by engineering dipole orientation.  Section 4 presents the main conclusions of the paper.

\section{Theoretical background}
\subsection{Theoretical foundation of 2.5D FEM approach}

In this section, we shall discuss how the 2.5D FEM approach is realized. We consider the  coupling between single emitter and an infinite long  waveguides. With the combination of Maxwell's equations and the constitutive relations, we have the following wave equation,
\begin{equation}\label{max1}
[\nabla  \times \frac{1}{{{{\bar{\bm\mu}_r}}}}\nabla  \times  - {\text{k}}_0^2{\bar{\bm\varepsilon} _r}({\bf{r}})]{\bf{E}}({\bf{r}},\omega ) - {\omega ^2}{\mu _0}{\bf{p}}({{\bf{r}}_0},\omega ) = 0,
\end{equation}
where ${\bar{\bm\mu}_r}$ and ${\bar{\bm\varepsilon}_r} ({\bf{r}})$ are relative permeability and relative dielectric constant respectively, ${{{k}}_0}$ is vacuum wave number, and ${\bf{p}}({{\bf{r}}_0},\omega )$ is the dipole moment of the emitter. We introduce a weighting function ${\bf{F}}({\bf{r}},\omega )$ to test  \Eq{max1}, and integrate over the entire modeling domain,
\begin{equation}\label{max3}
\begin{aligned}
L &= \iiint\limits_V {\left[ {\nabla  \times \frac{1}{{{\bar{\bm\mu}_r}}}\nabla  \times  - k_0^2{\bar{\bm\varepsilon} _r}({\bf{r}})} \right]}{\bf{E}}({\bf{r}},\omega ) \cdot {\bf{F}}({\bf{r}},\omega )dv - \iiint\limits_V {{\omega ^2}}{\mu _0}{\bf{p}}(\omega ,{{\bf{r}}_0}) \cdot {\bf{F}}({\bf{r}},\omega )dv\\
& = \iiint\limits_V {\frac{1}{{{\bar{\bm\mu}_r}}}(\nabla  \times }{\bf{E}}({\bf{r}},\omega )) \cdot (\nabla  \times {\bf{F}}({\bf{r}},\omega ))dv - \iiint\limits_V {k_0^2{\bar{\bm\varepsilon} _r}({\bf{r}})}{\bf{E}}({\bf{r}},\omega ) \cdot {\bf{F}}({\bf{r}},\omega )dv\\
& - \iiint\limits_V {{\omega ^2}}{\mu _0}{\bf{p}}(\omega ,{{\bf{r}}_0}) \cdot {\bf{F}}({\bf{r}},\omega )dv + \mathop{{\int\!\!\!\!\!\int}\mkern-21mu \bigcirc}\limits_{\partial V}
 {{\bf{F}}({\bf{r}},\omega ) \cdot \left[ {\frac{1}{{{\bar{\bm\mu}_r}}}{\bf{n}} \times \nabla  \times {\bf{E}}({\bf{r}},\omega )} \right]} ds,
\end{aligned}
\end{equation}
where $\partial V$ denotes the surface that encloses volume $V$, and ${\bf{n}}$ denotes the outward normal unit vector to the surface of the modeling domain. We want to find ${\bf{E}}({\bf{r}},\omega ) $
that minimizes Eq.~\ref{max3} over all test functions. However, Eq.~(\ref{max3}) is a triple integral where the length of $z$ axis is infinite, and it is a challenging problem for the domain discretization and construction of interpolation functions. Next, we will show how to decompose the 3D problem into a sequence of 2D problems by Fourier transforming of electric field ${\bf{E}}({\bf{r}})$ and dipole moment ${\bf{p}}({{\bf{r}}_0})$, and by selecting appropriate test function for each mode of the waveguide.

For short, ${\bf{p}}({{\bf{r}}_0},\omega )$, ${\bf{E}}({\bf{r}},\omega )$ and ${\bf{F}}({\bf{r}},\omega )$  will be denoted  as ${\bf{p}}({{\bf{r}}_0})$, ${\bf{E}}({\bf{r}})$ and ${\bf{F}}({\bf{r}})$ onwards. We introduce  Fourier expansion over the relevant quantities, i.e., ${\bf{E}}({\bf{r}}) = \frac{1}{{2\pi }}\int\limits_{ - \infty }^\infty  {{\bf{E}}(x,y,{k_{1z}})} {e^{ - i{k_{1z}}z}}d{k_{1z}}$ and ${\bf{p}}({{\bf{r}}_0}) = \frac{1}{{2\pi }}{\bf{p}}({x_0},{y_0})\int\limits_{ - \infty }^\infty  {{e^{ - i{k_{2z}}(z - {z_0})}}d{k_{2z}}} $, where ${\bf{E}}(x,y,{k_{1z}}) = \hat xE_x^{{k_{1z}}}(x,y,{k_{1z}}) + \hat yE_y^{{k_{1z}}}(x,y,{k_{1z}}) + \hat zE_z^{{k_{1z}}}(x,y,{k_{1z}})$.  Due to the invariance of the dielectric function along $z$ direction, i.e., $\varepsilon ({\bf{r}}) = \varepsilon (x,y)$,  the resulting formulation becomes a sequence of uncoupled 2D problems (2.5D problem) rather than a sequence of coupled 2D problems (amount to 3D problem). Substituting  ${\bf{E}}({\bf{r}})$ and ${\bf{p}}({{\bf{r}}_0})$ into Eq.~(\ref{max3}) and selecting a mono-modal test function of the form ${{\bf{F}}_n}({\bf{r}}) = {{\bf{F}}_n}(x,y,z) = {{\bf{F}}_n}(x,y,{k_{nz}}){e^{i{k_{nz}}z}}$, we can rewrite the weak formulation as
following
\begin{equation}\label{max7}\centering
\begin{aligned}
L &= \iint\limits_{X,Y} {dxdy\left\{ {\frac{1}{{{\bar{\bm\mu}_r}}}{\bf{CurlE}}(x,y,{k_{nz}}) \cdot {\bf{Curl}}{{\bf{F}}_n}(x,y,{k_{nz}}) - k_0^2{\bar{\bm\varepsilon} _r}({\bf{r}}){\bf{E}}(x,y,{k_{nz}}) \cdot {{\bf{F}}_n}(x,y,{k_{nz}})} \right\}}\\
 &- \iint\limits_{X,Y} {dxdy\left\{ {{\omega ^2}{\mu _0}{\bf{p}}({x_0},{y_0}) \cdot {{\bf{F}}_n}(x,y,{k_{nz}})} \right\}} + \mathop{{\int\!\!\!\!\!\int}\mkern-21mu \bigcirc}\limits_{\partial V}
 {{{\bf{F}}_n}({\bf{r}},\omega )}  \cdot \left[ {\frac{1}{{{\bar{\bm\mu}_r}}}{\bf{n}} \times \nabla  \times {\bf{E}}({\bf{r}},\omega )} \right]ds,
\end{aligned}
\end{equation}
where ${\bf{CurlE}}(x,y,{k_{nz}}) = \left[ {\hat x\left( {\frac{{\partial E_z^{{k_{nz}}}}}{{\partial y}} + i{k_{nz}}E_y^{{k_{nz}}}} \right) + \hat y\left( { - \frac{{\partial E_z^{{k_{nz}}}}}{{\partial x}} - i{k_{nz}}E_x^{{k_{nz}}}} \right) + \hat z\left( {\frac{{\partial E_y^{{k_{nz}}}}}{{\partial x}} - \frac{{\partial E_x^{{k_{nz}}}}}{{\partial y}}} \right)} \right]$. Test functions (${{\bf{F}}_n}$) are selected as Fourier modes, including the waveguide continuum, forming  a complete  and orthogonal basis in which any arbitrary field can be expanded. In Eq.~(\ref{max3}), such typical selection of test function allows us to obtain a trivial integration along $z$ axis without coupling Fourier modes. Subsequently, we obtain a series of decoupled planar integration in terms of ${\bf{E}}(x,y,{k_{nz}})$,  ${\bf{Cur}}{{\bf{l}}}{\bf{E}}(x,y,{k_{nz}})$, ${{\bf{F}}_n}(x,y,{k_{nz}})$ and ${{\bf{Cur}}}{{\bf{l}}}{{\bf{F}}_n}(x,y,{k_{nz}})$, which contain the same propagation constant ${k_{nz}}$. Notably, the 2D integration given by Eq.~(\ref{max7}) for each ${k_{nz}}$ can be solved as a standard 2D boundary-value problem by employing traditional finite element solution procedures, including discretization, selection of basis functions and assembling of the sparse matrix \cite{Jianming}. Coefficients ${k_{nz}}$ correspond to the Fourier frequencies in the $z$-direction. The number of such frequencies is often smaller than the number of basis funcions that one may select on the $z$-direction on a traditional finite element method, since higher-order methods like Fourier exhibits higher-order convergence. Additionally, the uncoupling among Fourier modes produces great computational savings with respect to traditional finite element methods, since it enables to decouple the original 3D problem into a sequence of uncoupled 2D problems, namely, one per Fourier mode. While a 3D problem can be interpreted from the algebraic point of view as a dense matrix of 2D problems, in the proposed approach, the resulting system becomes a diagonal matrix of 2D problems, which can be solved in a fraction of the time. In Eq.~(\ref{max7}), the dipole moment is invariant with respect to the propagating constant, which amounts to a  time-harmonic line current source independent of ${k_{nz}}$. Fourier frequencies ${k_{nz}}$ are selected in such a way that the source is well approximated in the frequency domain when using its Fourier series expansion.


\subsection{Rate of energy dissipation in inhomogeneous environment}
According to Poynting's theorem, the radiated power of any current distribution with a harmonic time dependence in a linear medium has to be identical to the rate of energy dissipation $P$ \cite{Cambridge,Snyder1983,Jackson1999}. If we introduce the dipole's current density such as in Eq.~\ref{max3}, i.e., ${\bf{J}}({\bf{r}},\omega ) =  - i\omega {\bf{p}}\delta ({\bf{r}} - {{\bf{r}}_{\text{0}}})$, we obtain the emitted power as
\begin{equation}\label{9}
P = \frac{{dW}}{{dt}} = \frac{\omega }{2}\operatorname{Im} \left\{ {{{\bf{p}}^ * } \cdot {\bf{E}}({{\bf{r}}_0})} \right\},
\end{equation}
where  ${\bf{E}}({{\bf{r}}_0})$ is evaluated at the dipole's origin ${{\bf{r}}_0}$. To formulate the spectrum presentation of the radiated power, we introduce differential emission rate $P(k_{nz})$, the integration of which over $k_{nz}$ gives the total emission rate $P$. We estimate the normalized differential energy dissipation rate ${P({k_{nz}})}/P_0$ for a particular $k_{nz}$, where ${P({k_{nz}})}$ is given by
 \beq
 P({k_{nz}}) = \frac{\omega }{2}\operatorname{Im} \left\{ {{{\bf{p}}^ * } \cdot {\bf{E}}({x_0},{y_0};{k_{nz}})} \right\},
 \eeq
 and ${P_0} = \int_{ - \infty }^ {+\infty}  {{P_0}({k_{nz}})} d{k_{nz}}$  is the emitted power by the same dipole source into all modes of a homogeneous material  where the emitter is seated.



The dyadic Green's function provides  us with an alternative approach to study spontaneous decay rate $\gamma $ of a two-level quantum system in an arbitrary optical environment \cite{Søndergaard2007,
quantum}. The differential spontaneous decay rate $\gamma ({k_{nz}})$  into each mode with propagating constant ${k_{nz}}$ can be normalized by the free-space spontaneous decay rate ${\gamma _0} = \frac{{\omega _0^3{{\left| p  \right|}^2}}}{{3\pi {\varepsilon _0}\hbar {c^3}}}$, and written in terms of spectrum expanding of Green's tensor as
\begin{equation}\label{gamma}
\frac{{\gamma }({k_{nz}})}{{{\gamma _0}}} = \frac{{6\pi c}}{{{\omega _0}}}\left[ {{\bf{n}}_p ^T \cdot \operatorname{Im} \left\{ {{\overline{G}} ({{\bf{r}}_0},{{\bf{r}}_0},\omega ;{k_{nz}})} \right\} \cdot {{\bf{n}}_p }} \right],
\end{equation}
with ${{\bf{n}}_p }$ being the unit vector in the direction of the dipole moment $\bf{p}$
, and ${\overline{G}} ({{\bf{r}}_0},{{\bf{r}}_0},\omega ; {k_{nz}})$ is the Green's tensor at the dipole position. The normalized differential spontaneous decay rate ${{\gamma ({k_{nz}})} \mathord{\left/
 {\vphantom {{\gamma ({k_{nz}})} {{\gamma _0}}}} \right.
 \kern-\nulldelimiterspace} {{\gamma _0}}}$ into each mode calculated by the dyadic Green's function method is equivalent to the normalized differential energy dissipation rate ${{P({k_{nz}})} \mathord{\left/
 {\vphantom {{P({k_{nz}})} {{P_0}}}} \right.
 \kern-\nulldelimiterspace} {{P_0}}}$ calculated by the 2.5 FEM approach, i.e., ${{\gamma ({k_{nz}})} \mathord{\left/
 {\vphantom {{\gamma ({k_{nz}})} {{\gamma _0}}}} \right.
 \kern-\nulldelimiterspace} {{\gamma _0}}} = {{P({k_{nz}})} \mathord{\left/
 {\vphantom {{P({k_{nz}})} {{P_0}}}} \right.
 \kern-\nulldelimiterspace} {{P_0}}}$. We  also  introduce the $\beta $ factor to describe the
fraction of the emitted energy that is coupled to the guided mode. This factor is defined as $\beta_m  = {{{P_m}} \mathord{\left/
 {\vphantom {{{P_m}} {{P_{total}}}}} \right.
 \kern-\nulldelimiterspace} {{P_{total}}}}$, where ${{P_m}}$ is the dissipation rate into the $m$ guided mode  obtained by intergrating $P(k_{nz})$ around the guided mode resonance, and ${{P_{total}}}$ is the total dissipation rate into all modes.


\section{Coupling between quantum emitter and infinite nanowire}

\subsection{Single mode cylindrical metallic nanowire}

\begin{figure}\centering
\setlength{\abovecaptionskip}{-2cm}
\includegraphics[scale=0.42]{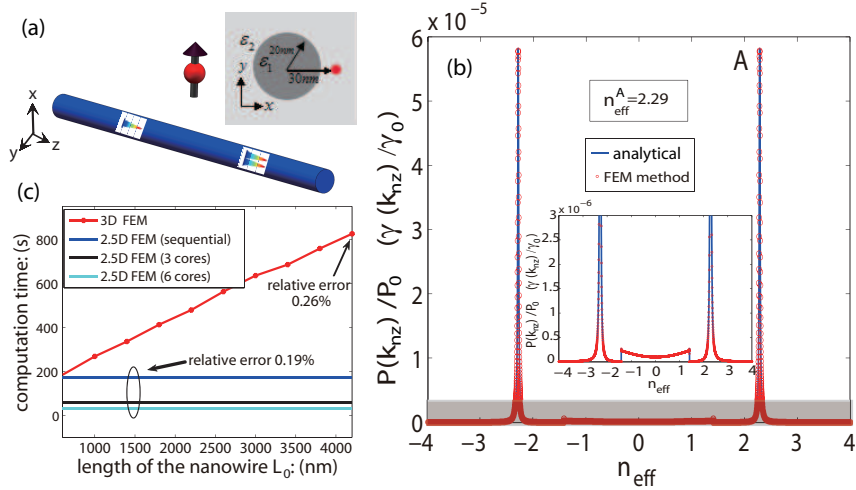}
\caption{\label{circle}   Spontaneous emission in a single nanowire. (a) Schematic of the coupling between a metallic nanowire and a quantum emitter oriented in the radial direction. (b) Normalized differential energy dissipation rate ${P({k_{nz}})}/{P_0}$ and normalized differential spontaneous emission rate ${\gamma({k_{nz}})}/{\gamma_0}$ as functions of the effective index of different modes. The normalized differential dissipation rate of the emitter coupled to each mode calculated by the 2.5D FEM  (red circle line), fully agree with the normalized differential emission rate calculated by the dyadic Green's function method (blue line). The insets of (b) is the enlarged view of the shadow region of (b). (c) Length dependence of computation time for the 3D FEM and for the 2.5D FEM when solving an infinite nanowire with different parallel computations.}
\end{figure}
To illustrate the validity of the 2.5 FEM approach, we compare our numerical results with analytical solutions obtained by using the Green's dyadic method. The coupling between waveguide  and emitter is sketched in Fig.~\ref{circle}(a), in which we consider a dipole (frequency ${f_0} = 300THz$) placed close to the surface of an infinitely long metallic nanowire ($R = 20nm$) coupled into the surface plasmons of the conducting waveguide. The wire with electric permittivity $\epsilon_{1}=-50+0.6j$ is surrounded by a material with $\epsilon_{2}=2$, and the dipole moment is oriented along the $x$ axis. The inset of Fig.~\ref{circle}(a) shows the profile of a 2D cross-section of the model.

The normalized differential emission rate with modal index ${n_{eff}}$, i.e., ${P({k_{nz}})}/P_0$, is calculated by our 2.5 FEM approach.  As can be seen from Fig.~\ref{circle}(b), our FEM result show an excellent agreement with that calculated by the dyadic Green's function method. Moreover,  there are three critical features shown in Fig.~\ref{circle}(b), which match extremely well when employing the two independent methods. First, there exist two peaks when the effective indexes is around $ \pm 2.29$, which are essentially the effective index of the fundamental mode  of the waveguide (${n_{eig}} = 2.29$). The two methods also provide the same maximal value of the peak for waveguide with the same losses, as Fig.~\ref{circle}(b) shows.
Second, when $\left| {{n_{eff}}} \right| > \sqrt {{\varepsilon _2}}  = \sqrt 2 $, the plasmons are evanescent in the radial direction, and the dissipation rate of the dipole rapidly declines. Taking into account the medium absorption, i.e., a small value of $\operatorname{Im} ({\varepsilon _1})$, the Dirac-delta function spectrum of the bound mode is approximately broadened into a Lorentz-like lineshape, as shown in the inset of Fig.~\ref{circle} (b), the half-width at half maximum (HWHM) is determined by $\operatorname{Im} ({\varepsilon _1})$. That is to say, in the evanescent region,  the main contribution to the dipole emission rate comes from excitation of plasmonic modes on the nanowire, which leads to an enhancement of the decay
rate of the emitter  and channeling of its emission into a single propagating plasmonic mode. Third, it is worth noticing, in the traveling-wave part ($\left| {{n_{eff}}} \right| < \sqrt {{\varepsilon _2}}  = \sqrt 2 $), the contribution of the emission rate comes from the freely propagating photons. Thus, we conclude that our 2.5 FEM approach is verified from the comparison with the independent Green's function technique. To compare the computational efficiencies of 2.5D FEM with its 3D counterpart, we plot Fig.~\ref{circle}(c) in which we show the length dependence of computation time for the 3D counterpart by using mode matching method. The 2.5D FEM approach can decompose a 3D problem into a series of decoupled 2D problems, which can be computed in parallel so the computational efficiency is significantly increased. According to the discussion of \cite{PRB}, the variation in the total decay rate is reduced by increasing the length of nanowire. The relative error at $L = 4200nm$ is about $0.26\% $, and the computing time is 825.3s, which is calculated by 3D FEM. However, with appropriate selections of ${k_{nz}}$ and its calculating ranges (we selected 228 ${k_{nz}}$ with the range of $[ - 5{k_0},5{k_0}]$), the relative error is about $0.19\% $, along with a calculating time of 172s calculated by 2.5D FEM in the sequential (no parallel) version, and 57s with 3 cores.  Thus, the 2.5D FEM indeed needs less computation time and provides higher precision compared with its 3D counterpart.

\begin{figure}
\setlength{\abovecaptionskip}{-1cm}
\centering
\includegraphics[scale=0.36]{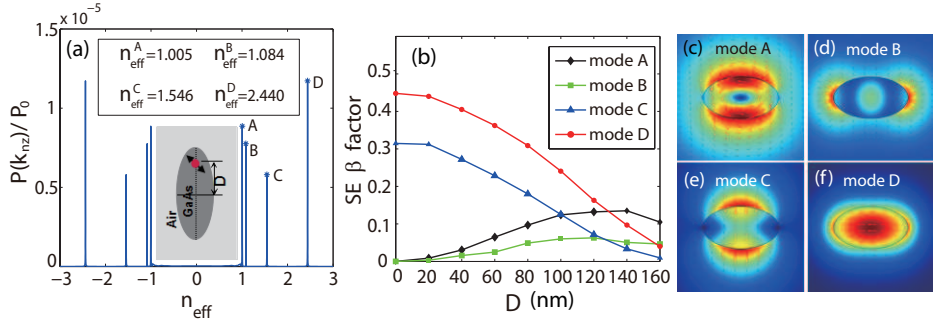}
\caption{\label{ellipse} SE in an infinite elliptical photonic nanowire. (a) Rate of differential energy dissipation as a function of the effective index. The four guided modes have been excited, i.e., modes A, B, C, and D. (b) The fraction of the emission coupled to each of the four guided eigen modes ($\beta$ factor) as a function of source distance $\ D$ from the center of the nanowire. (c)-(f) The electric field distributions of the four guided modes, where the arrows denote the field intensity and orientation.  }
\end{figure}

\subsection{Multi-mode elliptical nanowire}


%

As a known fact, it is difficult to study the emission of  single emitter to  elliptical  waveguides or others with  irregular cross section \cite{ellipse}. However, such problems  can be  easily handled by using our 2.5D FEM approach. In the following, we consider a semiconductor photonic nanowire made of GaAs (refractive index ${n_1} = 3.45$) and surrounded by an air cladding (${n_2} = 1$). The wire has an elliptical cross section, with major diameter $360nm$, twice the length of the minor diameter, and the emitter is placed on the long axis of a elliptical section, as the inset of  Fig.~\ref{ellipse}(a) shows.

We obtain the normalized differential dissipation rate of the dipole ${{P({k_{nz}})} \mathord{\left/
 {\vphantom {{P({k_{nz}})} {{P_0}}}} \right.
 \kern-\nulldelimiterspace} {{P_0}}}$ as a function of the different modes from Fig.~\ref{ellipse}(a). The four guided modes  are labelled as mode A, B, C, and D respectively.  We introduce a little material loss $\operatorname{Im} ({\varepsilon _1})$ to the wire to better visualize the emission peaks given by the four modes in our numerical calculation. We display the fraction of the emission ($\beta $ factor) coupled into each  guided mode as a function of the distance between dipole position and the center of ellipse section in Fig.~\ref{ellipse}(b). Figures~\ref{ellipse}(c)-\ref{ellipse}(f) discribe the electric field distribution of the four guided modes. As shown in Fig.~\ref{ellipse}(e), the electric field of mode C ($n_{eff}^3$) in the center of waveguide goes along the $y$ axis, which implies the  mode C can only be stimulated by a central dipole with $y$-component. Similarly, mode D ($n_{eff}^4$) can not be stimulated by a central dipole without $x$-component. Hence, in order to excite all the guided modes of the waveguide,  We set the emitter with two equal dipole components, i.e., $\bm p=\hat{x}p_0+\hat{y}p_0$. The electric field distributions of mode C and D show that the field intensity decrease gradually as the distance $D$ increase, as shown in Figs.~\ref{ellipse}(e)-\ref{ellipse}(f). Hence, the coupling between the dipole and the modes will decrease for large $D$, which is consistent  with the results shown on Fig.~\ref{ellipse}(b). The field distributions of the other two guided mode B and mode C indicate that field intensity increase to a its maximum as the distance $D$ increases and crosses an optimal distance, as illustrated in Figs.~\ref{ellipse}(c)-\ref{ellipse}(d), which  is also consistent with the tendency of coupling efficiency change shown in Fig.~\ref{ellipse}(b).

\begin{figure}\centering
\setlength{\abovecaptionskip}{-3cm}
\includegraphics[scale=0.50]{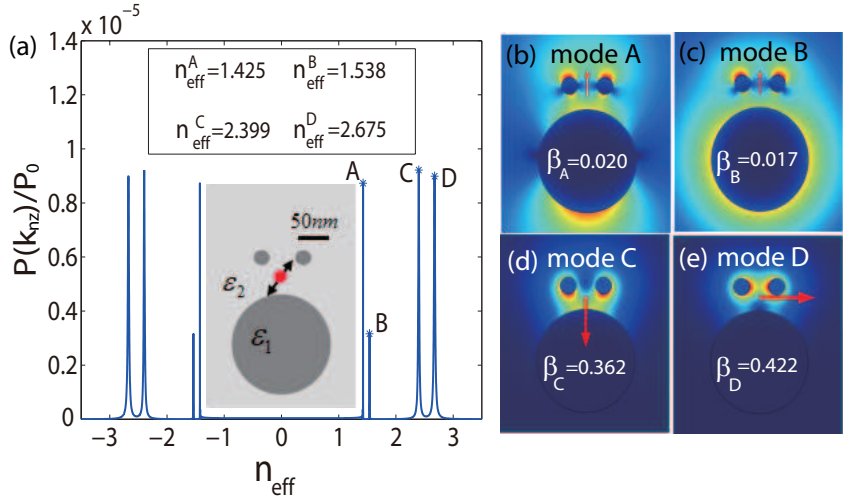}
\caption{\label{multicore} SE into a multi-core waveguide. (a) Differential energy dissipation rate of a single dipole relative to the homogeneous material (${\varepsilon _2} = 2$) dissipation rate as a function of effective index.  The guided modes propagating along the same direction have been excited, i.e., modes A, B, C, and D. (b)-(e) Electric field distributions of the four guided modes, where the arrows indicate the field of each guided mode in the dipole position. The $\beta$ factors of the four guided plasmonic modes are also shown within the profile of each mode. }
\end{figure}

\subsection{Multi-core nanowire}
We proceed to discuss the application of our 2.5D FEM approach to multi-core waveguide, which has the potential to expand the information capacity in our current optical communication network,  as discussed by Tu et al.\cite{Tu2012,Rasmussen2008}. We propose a multi-core plasmonic waveguide,  and place the emitter in the middle of two single-mode metallic nanowires (${r_1} = 20nm$)  and a multi-mode nanowire (${r_2} = 120nm$). The metallic cores with ${\varepsilon _1} =  - 50 + 0.6j$ are surrounded by a material with ${\varepsilon _2}  = 2$, and the dipole has two equal dipole moment components along $x$ and $y$ axis, as shown in the inset of Fig.~\ref{multicore}(a).

Figure.~\ref{multicore}(a) shows the normalized differential emission rate $P(k_{nz})/P_0$ as a function of the mode effective index of the waveguide. The multi-core waveguide supports four guided modes,  labeled by A, B, C, and D.  Figures~\ref{multicore}(b)-\ref{multicore}(e) show the electric filed distribution $|\bm E|$ corresponding to the four excited plasmonic modes, and the arrows
denote the intensity and direction of the electric filed at the dipole position. Obviously, the mode D with the maximum $|\bm E|$ at the dipole position indicates that the coupling efficiency of the dipole to the guided mode is highest, as confirmed by the maximal value of the $\beta$ factor ($\beta_D$). On the other hand, the minimum $|\bm E|$ of mode B indicates the lowest coupling efficiency, and corresponds to  the minimum value of the $\beta$ factor ($\beta_B$). The $\beta$ factors of Fig.~\ref{multicore}(b)-\ref{multicore}(e) also illustrate that the energy mainly dissipates into the four guided plasmonic modes rather than other evanescent waves or freely propagating photons. It is important to point out that the merit of the 2.5D FEM is that we can obtain the emission rate coupled into arbitrary modal channels for waveguides with arbitrary cross-sections, as well as obtain each mode coupling efficiency for arbitrary dipole position.


\begin{figure}\centering
\includegraphics[scale=0.40]{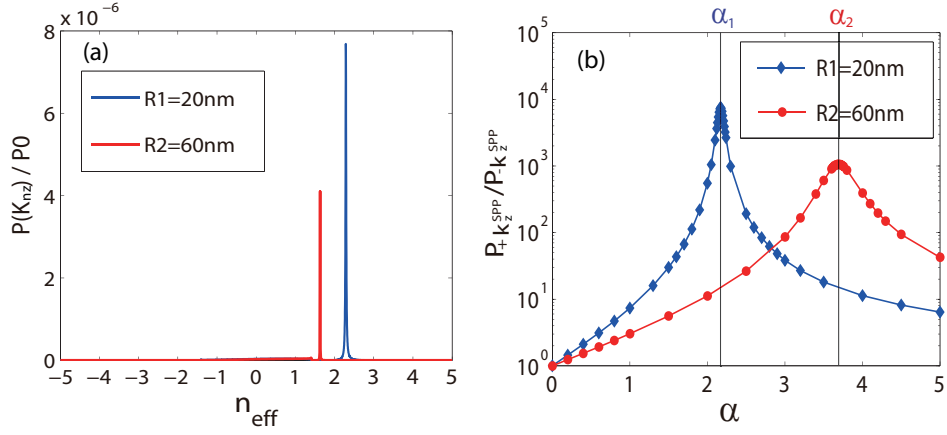}
\caption{\label{undirect}   Perfectly unidirectional excitation. The blue line and red line of (a) show the perfect unidirectional excitation of guided modes, in the case of a wire of radius ${R_1} = 20nm$ and
${R_2} = 60nm$ respectively. (b) The ${P_{ + k_z^{SPP}}}/{P_{ - k_z^{SPP}}}$ as a function of polarization of the dipole for the two case.  In each case, there is an optimal $\alpha $ where the mode with ${- k_z^{SPP}}$ suppressed completely.  }
\end{figure}

\subsection{Unidirectional excitation of electromagnetic guided mode}
In this section, we discuss the flexibility of dipole orientation in our 2.5D FEM approach and possible applications. As a concrete example, we study the unidirectional excitation of bound modes using rotating dipoles \cite{science,Bliokh2015,Feber2015,sollner2014,Petersen2014}.
For the non-rotating dipoles in the aforementioned discussions, waveguides support  guided modes that can be excited in pairs, which share the same  effective index ${n_{eff}} $ but propagate in opposite directions. Interestingly, it is found that rotating dipoles  can possibly lead to unidirectional excitation of waveguide mode, due to the intrinsic spin-momentum locking of waveguide mode discovered by Bliokh \cite{Bliokh2015}. Here, we study the emission of a rotating dipole coupled to a metallic nanowire,  where the dipole moment  is given by ${\bf{p}} = \left( {{p_x},{p_y},{p_z}} \right) = \left( {1,0,j\alpha } \right)$. In particular, we study two  cylindrical metallic nanowires with ${\varepsilon _1} =  - 50 + 0.6j$, which are surrounded by material with ${\varepsilon _2} = 2$. The radius of the two nanowires are ${R_1} = 20nm$ and ${R_2} = 60nm$ respectively.

Figure~\ref{undirect}(a) shows the  excitation of guided modes of the two different metallic  nanowires. Clearly, the guided mode with ${ + k_z^{SPP}}$ is excited  with a much lager amplitude than the one with ${ - k_z^{SPP}}$. Such unidirectional mode excitation can be explained by the interference  between two individual dipoles $p_x$ and  $p_z$. The emission rates for  ${ + k_z^{SPP}}$ and  ${ - k_z^{SPP}}$ are given by superposing the contributions from both $p_x$ and $p_z$ as follows,
\begin{equation}\label{10}
{P_{ + k_z^{SPP}}} = \frac{\omega }{2}\operatorname{Im} ({{\bf{p}}^ * } \cdot {{\bf{E}}^{dip}}({{\bf{r}}_{\bf{0}}}))=\frac{\omega }{2}\left[ {\operatorname{Im} (E_x^{dip}) + \alpha \operatorname{Re} (E_z^{dip})} \right],
\end{equation}
\begin{equation}\label{11}
{P_{ - k_z^{SPP}}} = \frac{\omega }{2}\operatorname{Im} ({{\bf{p}}^ * } \cdot {{\bf{E}}^{dip}}({{\bf{r}}_{\bf{0}}}))=\frac{\omega }{2}\left[ {\operatorname{Im} (E_x^{dip}) - \alpha \operatorname{Re} (E_z^{dip})} \right],
\end{equation}
where $P_{ + k_z^{SPP}}$ and $P_{ - k_z^{SPP}}$ represent differential emission rates of modes that propagate  toward positive $z$ and negative $z$ respectively, and ${{\bf{E}}^{dip}}$ is the electric field of the excited guided SPP mode at the dipole's origin. From \Eq{11}, one can easily find that perfectly unidirectional excitation occurs when   ${P_{ - k_z^{SPP}}} = 0$,  which leads to
\begin{equation}\label{alfa}
\alpha  = {{\operatorname{Re} (E_x^{eig})} \mathord{\left/
 {\vphantom {{\operatorname{Re} (E_x^{eig})} {\operatorname{Im} (E_z^{eig})}}} \right.
 \kern-\nulldelimiterspace} {\operatorname{Im} (E_z^{eig})}},
\end{equation}
where $E_x^{eig}$($E_z^{eig}$) is the $x$($z$)-component of  the electric filed of the eigen mode at the dipole position. To completely suppress  the excitation of ${ - k_z^{SPP}}$ mode, we select ${\alpha_1 } = 2.17$ (${\alpha_2 } = 3.70$) for  nanowire with raidus of $20nm$  ($60nm$).  Such simple analysis is confirmed by the $\alpha$-dependence study of  the ratio between two dissipation rates, i.e., ${P_{ + k_z^{SPP}}}/{P_{ - k_z^{SPP}}}$, as
 shown in Fig.~\ref{undirect}(b). Remarkably, the maximum value of ${P_{ + k_z^{SPP}}}/{P_{ - k_z^{SPP}}}$ is attained at ${\alpha _1} = 2.17$ (or ${\alpha _2} = 3.70$ for the $60nm$ case), which is the exact value predicted by Eq.~(\ref{alfa}). Thus, for given parameters of the waveguide, there is an optimal $\alpha $ that ensures a maximal ${P_{ + k_z^{SPP}}}/{P_{ - k_z^{SPP}}}$, up to $10^4$ in our case, resulting in a perfect unidirectional excitation.


\section{Conclusions}
In conclusion, we proposed a 2.5D FEM approach to decompose a 3D coupled waveguide-emitter problem into a series of independent 2D problems, which can handle the excitation of arbitrary orientation sources for waveguides with arbitrary cross-section.  Considering the translation symmetry of waveguide structures, we successfully realized dimension reduction using Fourier transform and proper selection of test functions. To illustrate the validity of the 2.5D FEM approach, we have calculated the dipole emission rate coupled into each mode for various waveguide structure. We compared  our method with the dyadic Green's function method for SE in a single mode cylindrical metallic nanowire. The comparison confirms that our 2.5D FEM results coincide with those obtained by the dyadic Green's approach for the SE in a single mode cylindrical metallic nanowire and our approach is more effective for other complex waveguides. We derived explicit expressions for the coupling strength of an dipole to the guided modes and discussed  how it depends on the dipole's position and wire parameters. To reveal flexibility of the 2.5D FEM model, we have also applied it to a multi-mode elliptical waveguide and a multi-core waveguide. We further applied the method to the unidirectional excitation of guided modes via a rotating dipole. Depending upon the polarization state of the dipole, the decay rate of the forward propagating mode or inverse propagating mode is enhanced or suppressed. We also concluded that for a given waveguide there exists an optimal polarization state to achieve perfect unidirectional excitation of guided modes.

As an outlook, the 2.5D FEM can be extended to waveguide with finite length. In this scenario, the modes with ${k_{nz}}$ will bounce back and forth between the two terminations of the waveguide. This extension may be useful to study finite structures including cavities and coupled waveguide-cavity systems.

\section*{Acknowledgment}
This work was supported by the National Natural Science Foundation of China (Grant No. 61405067) and Foundation for Innovative Research Groups of the Natural Science Foundation of Hubei Province (Grant No. 2014CFA004).


\begin{thebibliography}{10}
\bibitem{Lodahl2015} P. Lodahl, S. Mahmoodian, and S. Stobbe, ``Interfacing single photons and single quantum dots with photonic nanostructures,'' Rev. Mod. Phys. {\bf 87,} 347--400 (2015).

\bibitem{Frimmer2011} M. Frimmer, Y. Chen, and A. F. Koenderink, ``Scanning emitter lifetime imaging microscopy for spontaneous emission control,'' Phys. Rev. Lett. {\bf 107,} 123602 (2011).

\bibitem{Lund2008} T. Lund-Hansen, S. Stobbe, B. Julsgaard, H. Thyrrestrup, T. S\"{u}ünner, M. Kamp, A. Forchel, and P. Lodahl, ``Experimental realization of highly efficient broadband coupling of single quantum dots to a photonic crystal waveguide,'' Phys. Rev. Lett. {\bf 101,} 113903 (2008).

 \bibitem{tonglimin2013} X. Guo, Y. Ma, Y. Wang, and L. Tong, ``Nanowire plasmonic waveguides, circuits and devices,'' Laser Photon. Rev. {\bf 7,} 855--881 (2013).

\bibitem{Femius2006} A. F. Koenderink, M. Kafesaki, C. M. Soukoulis, and V. Sandoghdar, ``Spontaneous emission rates of dipoles in photonic crystal membranes,''  J. Opt. Soc. Am. B {\bf 23,} 1196--1206 (2006).

\bibitem{Hummer2013} T. H\"{u}ümmer, F. J. Garc\'{i}ía-Vidal, L. Mart\'{i}ín-Moreno, and D. Zueco, ``Weak and strong coupling regimes in plasmonic QED,'' Phys. Rev. B {\bf 87,} 115419 (2013).

\bibitem{gongqihuang2014} Y. Bian  and Q. Gong, `` Metallic nanowire-loaded plasmonic slot waveguide for highly confined light transport at telecom wavelength,''  IEEE J. Quantum Electron. {\bf 49,} 870--876 (2013).

\bibitem{Kneipp1997}  K. Kneipp, Y. Wang,  H. Kneipp, L. T. Perelman, I. Itzkan, R. Dasari,  M. S. Feld, ``Single molecule detection using surface-enhanced Raman scattering (SERS),'' Phys.  Rev. Lett. {\bf 78,} 1667--1670 (1997).


\bibitem{Chen2009}  X. W. Chen, V. Sandoghdar, and M. Agio,  ``Highly efficient interfacing of guided plasmons and photons in nanowires,'' Nano Lett.  {\bf 9,} 3756--3761 (2009).

\bibitem{Ulrik2013OL} S. Kumar, A. Huck, Y. W. Lu, and U. L. Andersen, ``Coupling of single quantum emitters to plasmons propagating on mechanically etched wires,'' Opt. Lett. {\bf 38,}  3838--3841 (2013).

\bibitem{Søndergaard2001} T. S\oøndergaard and B. Tromborg, ``General theory for spontaneous emission in active dielectric microstructures: Example of a fiber amplifier,'' Phys. Rev. A {\bf 64,} 033812 (2001).

\bibitem{Falk2009} A. L. Falk, F. H. L. Koppens, C. L. Yu, K. Kang, N. D. Snapp, A. V. Akimov, M. H. Jo, M. D. Lukin, and H. Park, ``Near-field electrical detection of optical plasmons and single-plasmon sources,'' Nat. Phys. {\bf 5,} 475--479 (2009).

\bibitem{Crozier2003}  K. B. Crozier, A. Sundaramurthy, G. S. Kino, and C. F. Quate, ``Optical antennas: resonators for local field  enhancement,'' J. Appl. Phys. {\bf 94,} 4632--4642 (2003).

\bibitem{Christian2002} C. Hermann and O. Hess, ``Modified spontaneous-emission rate in an inverted-opal structure with complete photonic bandgap,''  J. Opt. Soc. Am. B {\bf 19,} 3013--3018 (2002).





 \bibitem{Vos2004} P. Lodahl, A. F. van Driel, I. S. Nikolaev, A. Irman, K. Overgaag, D. Vanmaekelbergh, and W. L. Vos, ``Controlling the dynamics of spontaneous emission from quantum dots by photonic crystals,'' Nature (London) {\bf 430,} 654--657 (2004).

 \bibitem{Colas2010} G. C. des Francs, P. Bramant, J. Grandidier, A. Bouhelier, J. C. Weeber, and A. Dereux,  ``Optical gain, spontaneous and stimulated emission of surface plasmon polaritons in confined plasmonic waveguide,'' Opt. Express {\bf 18,}  16327--16334 (2010).


\bibitem{Chang2007} D. E. Chang, A. S. S\oørensen, P. R. Hemmer, and M. D. Lukin, ``Strong coupling of single emitters to surface plasmons,'' Phys. Rev. B {\bf 76,} 035420 (2007).


\bibitem{Hecker1999} N. E. Hecker, R. A. Hopfel, N. Sawaki, T. Maier, and G. Strasser, ``Surface plasmon enhanced photoluminescence from a single quantum well,'' Appl. Phys. Lett. {\bf 75,} 1577--1579 (1999).





\bibitem{perter2008} J. Johansen, S. Stobbe, I. S. Nikolaev, T. Lund-Hansen, P. T. Kristensen, J. M. Hvam, W. L. Vos, and P. Lodahl, ``Size dependence of the wavefunction of self-assembled InAs quantum dots from time-resolved optical measurements,'' Phys. Rev. B {\bf 77,} 073303 (2008).


\bibitem{V.sandoghdar2006}  S. K\"{u}hn, U. Hakanson, L. Rogobete, and V. Sandoghdar, ``Enhancement of single molecule fluorescence using a gold nanoparticle as an optical nano-antenna,'' Phys. Rev. Lett. {\bf 97,} 017402 (2006).

\bibitem{Ulrik2013} S. Kumar, A. Huck, Y. T. Chen, and U. L. Andersen, ``Coupling of a single quantum emitter to end-to-end aligned silver nanowires ,'' Appl. Phys. Lett. {\bf 102,} 103106 (2013).


\bibitem{Chang2006} D.  E. Chang, A.  S. S\oørensen, P.  R. Hemmer,  and M.  D. Lukin, ``Quantum optics with surface plasmons,'' Phys. Rev. Lett. {\bf 97,} 053002 (2006).

\bibitem{Arcari2014} M. Arcari, I. S\"{o}öllner, A. Javadi, S. L. Hansen, S. Mahmoodian, J. Liu, H. Thyrrestrup, E.  H. Lee, J.  D. Song, S. Stobbe, and P. Lodahl, ``Near-unity coupling efficiency of a quantum emitter to a photonic crystal waveguide,'' Phys. Rev. Lett. {\bf 113,} 093603 (2014).




\bibitem{PRB} Y. T. Chen, T. R. Nielsen, N. Gregersen, P. Lodahl, and J. M\oørk, ``Finite-element modeling of spontaneous emission of a quantum emitter at nanoscale proximity to plasmonic waveguides,'' Phys. Rev. B {\bf 81,} 125431 (2010).

\bibitem{Niles2009} N. Gregersen and J. M\oørk, ``An improved perfectly matched layer for the eigenmode
expansion technique,'' Opt. Quantum Electron.  {\bf 40,} 957--966 (2009).

\bibitem{Chen2010} Y. Chen, N. Gregersen, T. R. Nielsen, J. M\oørk, and P. Lodahl, ``Spontaneous decay of a single quantum dot coupled to a metallic slot waveguide in the presence of leaky plasmonic modes,'' Opt. Express {\bf 18,} 12489--12498 (2010).

\bibitem{Pardo} D. Pardo, L. Demkowicz, C. Torres-Verd\'{i}ín, and C. Michler, ``PML enhanced with a self-adaptive goal-oriented hp-finite element method: simulation of through-casing borehole resistivity measurements,'' SIAM  J. Sci. Comput. {\bf 30,} 2948--2964 (2008).


\bibitem{Jakob2013} J. R. de Lasson, J. M\oørk, and P. T. Kristensen, ``Three-dimensional integral equation approach to light scattering, extinction cross sections, local density of states, and quasi-normal modes,'' J. Opt. Soc. Am. B {\bf 30,}  1996--2007 (2013).




\bibitem{ellipse} M. Munsch, J. Claudon, J. Bleuse, N. S. Malik, E. Dupuy, J.-M. G\'{e}érard, Y. Chen, N. Gregersen, and J. M\oørk, ``Linearly polarized, single-mode spontaneous emission in a photonic nanowire,'' Phys. Rev. Lett. {\bf 108,} 077405 (2012).


\bibitem{Jianming}  J. M. Jin, {\it The Finite Element Method in Electromagnetics} (Wiley, 1993).



\bibitem{Cambridge}  L. Novotny and B. Hecht, {\it Principles of Nano-optics} (Cambridge University, 2006).

\bibitem{Snyder1983}  A. W. Snyder and J. Love, {\it Optical Waveguide Theory } (Springer, New York, 1983).

\bibitem{Jackson1999}  J. Jackson, {\it Classical Electrodynamics} 3rd ed.  (Wiley, New York, 1999).


\bibitem{Søndergaard2007} T. S\oøndergaard, ``Modeling of plasmonic nanostructures: Green's function integral equation methods,'' Phys. Status Solidi B {\bf 244} (10){\bf ,}  3448--3462 (2007).

\bibitem{quantum} D. Dzsotjan, A. S. S\oørensen, and M. Fleischhauer, ``Quantum emitters coupled to surface plasmons of a nanowire: A Green's function approach,'' Phys. Rev. B {\bf 82,}  075427 (2010).





\bibitem{Tu2012} J. Tu, K. Saitoh, M. Koshiba, K. Takenaga, and S. Matsuo,  ``Design and analysis of large-effective-area heterogeneous trench-assisted multi-core fiber,''  Opt. Express, {\bf 20,} 15157--15170 (2012).

\bibitem{Rasmussen2008} P. D. Rasmussen, A. A. Sukhorukov, D. N. Neshev, W. Krolikowski, O. Bang, J. Lægsgaard, and Y. S. Kivshar, ``Spatiotemporal control of light by Bloch-mode dispersion in multi-core fibers,''  Opt. Express {\bf 16}  5878--5891 (2008).

\bibitem{science}  F. J. Rodr\'{i}guez-Fortu\~{n}o, G. Marino, P. Ginzburg, D. O'Connor, A.  Mart\'{i}nez, G. A. Wurtz, and A. V. Zayats, ``Near-field interference for the unidirectional excitation of electromagnetic guided modes,''  Science {\bf 340,} 328--330 (2013).

\bibitem{Bliokh2015} K. Y. Bliokh, D. Smirnova,  F. Nori, ``Quantum spin Hall effect of light,'' Science {\bf 348,}  1448--1451 (2015).

\bibitem{Feber2015} B. le Feber,	N. Rotenberg	and L. Kuipers, ``Nanophotonic control of circular dipole emission,'' Nat. Commun. {\bf 6,} 6695 (2015).

\bibitem{sollner2014} I. S\"{o}öllner, S. Mahmoodian, S. L. Hansen, L. Midolo, A. Javadi, G. Kir\u{s}šansk\.{e}ė, T. Pregnolato, H. El-Ella, E. H. Lee, J. D. Song, S. Stobbe, P. Lodahl, ``Deterministic photon-emitter coupling in chiral photonic circuits,'' Nature Nanotechnology {\bf 10,} 775--778 (2015).

\bibitem{Petersen2014} J. Petersen, J. Volz, A. Rauschenbeutel, ``Chiral nanophotonic waveguide interface based on spin-orbit interaction of light,'' Science {\bf 346,} 67--71 (2014).



\end{thebibliography}
\end{document}